\documentclass[12pt]{article}
\usepackage{amsmath,amsfonts}
\usepackage{amssymb}
\usepackage{amsthm}
\usepackage{times,txfonts}

\newtheorem{thm}{Theorem}[section]

\newtheorem{lemma}[thm]{Lemma}

\newtheorem{remark}[thm]{\it Remark}

\numberwithin{equation}{section}

\def\pf{\noindent{\it Proof.} \ }

\def\qed{\hfill $\square$}

\title{A geometric approach to tau-functions
of difference Painlev\'e equations}

\author{Teruhisa TSUDA   \\
Faculty of Mathematics, Kyushu University, \\  
Hakozaki, Fukuoka 812-8581, Japan.  \\
tudateru@math.kyushu-u.ac.jp}
\date{April 7, 2008}

\begin{document}
\maketitle
\begin{abstract}
We present  
a unified description of 
birational representation of  Weyl groups  
associated with T-shaped Dynkin diagrams,
by using a particular configuration of points  in the projective plane.
A geometric formulation of tau-functions is given in terms of 
defining polynomials of certain curves.
If the Dynkin diagram is of affine type ($E_6^{(1)}$, $E_7^{(1)}$  or $E_8^{(1)}$),
our representation gives rise to the difference Painlev\'e equations.
\end{abstract}

\renewcommand{\thefootnote}{\fnsymbol{footnote}}

\section*{Introduction}

\begin{figure}[h]
\begin{center}
\begin{picture}(200,85)

\put(40,17){$\underbrace{ \qquad \qquad \quad}$}
\put(66,0){$\ell_1$}

\put(101,17){$\underbrace{ \qquad \qquad \quad}$}
\put(128,0){$\ell_2$}

\put(90,48){$\left\{ \begin{array}{l} \\ \\ \\ \\ \end{array} \right.$}
\put(80,48){$\ell_3$}

\thicklines

\put(40,20){\circle{4}}  
\put(42,20){\line(1,0){8}}
\put(53,20){$\ldots$}
\put(78,20){\line(-1,0){8}}
\put(80,20){\circle{4}}    
\put(98,20){\line(-1,0){16}}
\put(100,20){\circle{4}}   
\put(102,20){\line(1,0){16}}
\put(120,20){\circle{4}}
\put(122,20){\line(1,0){8}}
\put(133,20){$\ldots$}
\put(158,20){\line(-1,0){8}}
\put(160,20){\circle{4}}   

\put(100,22){\line(0,1){16}}
\put(100,40){\circle{4}}   
\put(100,42){\line(0,1){8}}
\put(98,54){$\vdots$}
\put(100,66){\line(0,1){8}}
\put(100,76){\circle{4}}  


\end{picture}
\end{center}
\caption{Dynkin diagram $T_{\ell_1,\ell_2,\ell_3}$}  \label{fig:T}
\end{figure}
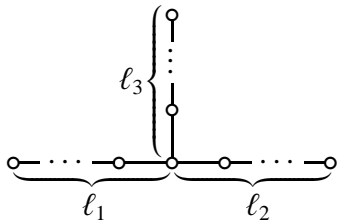

It is classically well-known that a certain birational representation of Weyl groups 
arises from point configurations. 
Let $X_{m,n}$ be the 
configuration space of $n$ points of
the projective space
${\mathbb P}^{m-1}$
in {\it general} position.
Then, the Weyl group 
corresponding to the Dynkin diagram 
$T_{2,m,n-m}$
(see Figure~\ref{fig:T})
acts birationally on $X_{m,n}$
and is generated by permutations of $n$ points
and the standard Cremona transformation with respect to each $m$ points;
see \cite{cob, do, kmnoy06}.

We shall put our attention to the two-dimensional case, that is, $m=3$ case.
Only if $n=9$, the affine case  occurs and the corresponding diagram reads $T_{2,3,6}=E_{8}^{(1)}$.
The lattice part of the affine Weyl group $W(E_{8}^{(1)})$ provides an interesting discrete dynamical system, 
called the
 {\it elliptic-difference Painlev\'e equation} \cite{org01,sak01}.
 This $(m,n)=(3,9)$ case 
was explored by Sakai \cite{sak01} 
(cf. \cite{oka})
in order to clarify the geometric nature of 
the affine Weyl group symmetry of Painlev\'e equations;
moreover he classified all the 
degeneration of the nine-points configuration in ${\mathbb P}^{2}$,
and as a result he completed the whole list of (second-order) discrete Painlev\'e equations.
In this context
the top of all the discrete and continuous Painlev\'e
equations is the elliptic-difference one, 
from which every other can be obtained through an appropriate limiting procedure.
The discrete Painlev\'e equations  
are divided into three types:
difference, $q$-difference or elliptic-difference one,
according to their corresponding rational surfaces 
(obtained by blowing up nine points of ${\mathbb P}^2$); see \cite{sak01}.
On the other hand, even in the two-dimensional case ($m=3$),
by considering certain particular configurations of point sets that are not only nine points,
one can enjoy Weyl groups associated to more various Dynkin diagrams
 \cite{loo, tsu06};
 see also \cite{tt} in the higher dimensional case.
\\

In this paper, 
we present a unified description of 
birational representation of  Weyl groups  
associated with T-shaped Dynkin diagrams (see Figure~\ref{fig:T}),
arising from a particular configuration of points  in the projective plane.
Our construction, 
in affine case: 
$E_6^{(1)}$, $E_7^{(1)}$ and $E_8^{(1)}$,
is relevant to the {\it difference} Painlev\'e
equations;
we refer to \cite{tsu06}
in the case of $q$-difference ones.

In Section~\ref{sect:bir}, 
we first start from $\ell_1+\ell_2+\ell_3$ points in ${\mathbb P}^2$
restricted on three lines $L_i$ $(i=1,2,3)$
meeting at a single point, 
where the $\ell_i$ points lie on each $L_i$. 
Let $X$ be the rational surface obtained from ${\mathbb P}^{2}$
by blowing up the $\ell_1+\ell_2+\ell_3$ points.
We can find naturally the root lattice of type $T_{\ell_1,\ell_2,\ell_3}$ included in 
the Picard group ${\rm Pic} (X)$ of the surface.
The corresponding Weyl group $W=W(T_{\ell_1,\ell_2,\ell_3})$
acts linearly on ${\rm Pic} (X)$. 
Next,
in order to lift this linear action $W:{\rm Pic} (X) \circlearrowleft$
to the level of birational transformations on the surface $X$ itself,
we introduce a geometric formulation of {\it tau-functions}.
Recall that a smooth rational curve 
with self-intersection $-1$
is said to be 
an {\it exceptional curve} (of the first kind);
see e.g.  \cite{bhpv}.
An element of $W$ induces a permutation among exceptional
curves on $X$, 
as analogous to 
the classical subject: 27 lines on a cubic surface and a Weyl group of type $E_6$.
The tau-function is defined by means of appropriately {\it normalized} defining functions of  the exceptional curves;
see (\ref{eq:nor}) and (\ref{eq:deftau}).
Today the notion of tau-functions is universal  
in the field of integrable systems
and it is characterized in several directions;
however, 
such an algebro-geometric idea of tau-functions 
first appeared in the study of the elliptic-difference Painlev\'e equation
by Kajiwara et al. \cite{kmnoy03}
(see also \cite{kmnoy06, tsu06, tt} for subsequent development).
By imposing on the tau-function 
a certain compatibility 
with the linear action $W:{\rm Pic} (X) \circlearrowleft$,
we finally obtain a birational representation of $W$ acting on 
the tau-function
and the {\it homogeneous} coordinates of 
${\mathbb P}^2$ (Theorem~\ref{thm:birat}).
Section~\ref{sect:aff} concerns the affine case
($E_6^{(1)}=T_{3,3,3}$, 
$E_7^{(1)}=T_{4,4,2}$ and
$E_8^{(1)}=T_{6,3,2}$); 
we demonstrate how to derive the difference Painlev\'e equation for each.

\section{Birational representation of Weyl groups and tau-functions}
\label{sect:bir}

We begin with considering the $\ell_1+\ell_2+\ell_3$ points in the complex projective plane ${\mathbb P}^2$
restricted on three lines 
$L_i$ $(i=1,2,3)$
meeting at a single point $P_0$, where we arrange $\ell_i$ points on each $L_i \setminus P_0$. 
Let ${\boldsymbol x}=[x_1:x_2:x_3]$ denote
the homogeneous coordinates of 
${\mathbb P}^2$.
By ${\rm PGL}_3(\mathbb C)$-action, 
we can  
normalize, without loss of generality, 
the three lines  and
the point configuration 
$\{  P_i^{m}\}_{1 \leq i \leq \ell_m; m=1,2,3}$
to be as follows:
\begin{align*}
L_i&=\{x_j=x_k\} \quad \text{for} \quad \{i,j,k\}=\{1,2,3\},  \\
P_i^{1}&=[b_i^1+1:b_i^{1}:b_i^{1}] \quad (1 \leq i \leq \ell_1), \\
P_i^{2}&=[ b_i^{2}  :b_i^2+1:b_i^{2}   ] \quad (1 \leq i \leq \ell_2), \\
P_i^{3}&=[b_i^{3}  :b_i^{3}  :b_i^3+1] \quad (1 \leq i \leq \ell_3),
\end{align*}
where
$3 b_1^m=-1+a_0+c_1+c_2+c_3-3c_m$
and
$b_{i+1}^m= b_i^m+ a_{i}^m$ $(1 \leq i \leq \ell_{m}-1; m=1,2,3)$ 
with
$c_m=\sum_{i=1}^{\ell_m-1} (1-\frac{i}{\ell_m})a_i^m$.
Here 
${\boldsymbol a}=(a_0, (a_i^m)_{1 \leq i \leq \ell_m-1; m=1,2,3} ) \in {\mathbb C}^{\ell_1+\ell_2+\ell_3-2}$
play roles of  free parameters.
We will show later the meaning of ${\boldsymbol a}$
as root variables corresponding to the Dynkin diagram $T_{\ell_1,\ell_2,\ell_3}$.

Let $\pi:X \to {\mathbb P}^2$ be the blow-up at the 
$\ell_1+\ell_2+\ell_3$ points $P^{m}_i$.
The {\it Picard group} of the surface $X$ thus obtained is
expressed as
\[
{\rm Pic} (X)
= {\mathbb Z} h \oplus 
\bigoplus_{
\begin{subarray}{l}
i=1,\ldots,\ell_m;\\ 
m=1,2,3
\end{subarray}
}
{\mathbb Z}  e_i^{m}
\quad
\left(\cong {\mathbb Z}^{\ell_1+\ell_2+\ell_3+1}\right).
\]
Note that in this case ${\rm Pic} (X)$ is isomorphic to the second cohomology group 
$H^2(X, {\mathbb Z})$
because $X$ is a rational surface.
Here we denote by $h$ the linear equivalent class of $\pi^{-1}$ of a line
and by  $e_i^{m}$ the class of exceptional curve $\pi^{-1}(P_i^{m})$.
The intersection form
$(\, | \,):{\rm Pic}(X)\times{\rm Pic}(X)\to {\mathbb Z}$
is given by
\[
(h|h)=1, \quad  ( e_i^{m} | e_j^{n} ) =-\delta_{i,j}\delta_{m,n}, \quad
( h  | e_i^{m} ) = 0.
\]

The anti-canonical class $-K_X=3 h -\sum_{i,m} e_i^{m}$
can be decomposed 
as $-K_X=D_1+D_2+D_3$,
where the classes
$
D_m=
h 
-\sum_{i=1}^{\ell_m}
e_i^{m}$ 
are represented
by (the proper transforms of) the three lines $L_m$ $(m=1,2,3)$.
Let $Q$ be the orthogonal complement of 
$\{ D_1, D_2,D_3 \}$ 
with respect to the intersection form
in ${\rm Pic} (X)$.
Then we see that
$Q$ is a root lattice generated by the $(-2)$-vectors  
$\alpha_{ij}^{m}=e_i^{m}-e_j^{m}$
and  
$\alpha_{ijk}=h-e_i^{1}-e_j^{2}-e_k^{3}$.
Moreover  we can choose a basis $B=\{\alpha_0=\alpha_{111}, 
\
\alpha_i^{m}=\alpha_{i,i+1}^{m}  
\ 
(1 \leq  i  \leq \ell_{m}-1, m=1,2,3) \}$
of $Q$.
The intersection graph (Dynkin diagram) of $B$ 
is of type $T_{\ell_1,\ell_2,\ell_3}$ and
looks as follows{\rm:}
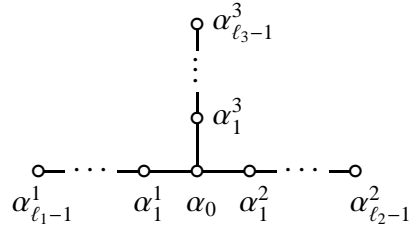
\begin{center}
\begin{picture}(200,85)

\thicklines

\put(40,20){\circle{4}}    \put(30,5){$\alpha_{\ell_1-1}^1$}
\put(42,20){\line(1,0){8}}
\put(53,20){$\ldots$}
\put(78,20){\line(-1,0){8}}
\put(80,20){\circle{4}}    \put(76,5){$\alpha_1^1$}
\put(98,20){\line(-1,0){16}}
\put(100,20){\circle{4}}    \put(96,5){$\alpha_0$}
\put(102,20){\line(1,0){16}}
\put(120,20){\circle{4}}    \put(116,5){$\alpha_1^2$}
\put(122,20){\line(1,0){8}}
\put(133,20){$\ldots$}
\put(158,20){\line(-1,0){8}}
\put(160,20){\circle{4}}    \put(158,5){$\alpha_{\ell_2-1}^2$}

\put(100,22){\line(0,1){16}}
\put(100,40){\circle{4}}   \put(106,38){$\alpha_{1}^3$}
\put(100,42){\line(0,1){8}}
\put(98,54){$\vdots$}
\put(100,66){\line(0,1){8}}
\put(100,76){\circle{4}}  \put(106,74){$\alpha_{\ell_3-1}^3$}

\end{picture}
\end{center}
We define 
the action of the simple reflection 
corresponding to a root 
$\alpha \in Q$  (i.e., $\alpha^2=-2$)
by
\[
R_\alpha(v)=v+(v|\alpha)\alpha
\]
for
$v \in {\rm Pic} (X)$.
Note that the intersection form is preserved by $R_\alpha$.
We also prepare the notations
$s_0=R_{\alpha_0}$ and 
$s_i^m=R_{\alpha_i^m}$,
for convenience.
We summarize below
the linear action of generators $s_0$ and $s_i^m$ on the basis of 
${\rm Pic} (X)$:
\begin{align}
s_0(h)&=2h-e_1^1-e_1^2-e_1^3, \nonumber
 \\
s_0(e_1^k)&=h-e_1^i-e_1^j \quad \text{for} \quad\{i,j,k\}=\{1,2,3\},
\\
s_i^m(e_{\{i,i+1\}}^m) &=e_{\{i+1,i\}}^m.
\nonumber
\end{align}
It is easy to check that
the {\it Weyl group} 
$W=\{ R_\alpha \}_{\alpha \in Q} 
= \langle s_0, s_i^m \ 
(1 \leq  i  \leq \ell_{m}-1, m=1,2,3)\rangle$ acting on
${\rm Pic}(X)$
indeed satisfies the fundamental relations \cite{kac}
specified by the Dynkin diagram 
$T_{\ell_1,\ell_2,\ell_3}$.

Now we shall extend the above linear action of $W : {\rm Pic}(X)\circlearrowleft$ 
to the level of birational transformations on the surface $X$.
We first fix the action on the  root variables
$a_0$ and $a_i^{m}$ $(1  \leq i \leq \ell_m-1, m=1,2,3)$
as 
\begin{equation}  \label{eq:act-on-a}
\begin{array}{ll}
s_0(a_0)=-a_0, 
&s_0(a_1^{m})=a_0 +a_1^{m},
\\
s_i^{m}(a_i^{m})=-a_i^{m}, 
&
s_i^{m}(a_{i \pm 1}^{m})=a_i^{m}+a_{i\pm 1}^{m} ,
\end{array}
\end{equation}
where $a_{0}^{m}=a_0$.
We consider a  sub-lattice $M=\bigsqcup_{n=1,2,3} M_n$
of ${\rm Pic}(X)$,
where 
\begin{align*}
M_n &\stackrel{\rm def}{=} W. \{  e_1^n\}
\\
&=
\left\{\Lambda =d h - \sum_{i,j} \mu^i_j e^i_j \in {\rm Pic} (X) \,  \bigg| 
\begin{array}{l} 
\Lambda^2=-1, \  (\Lambda | D_n)=1, \  (\Lambda | D_m)=0 \  (m\neq n)
\\
\text{$d \geq 0$; if $d>0$, then $\mu^i_j \geq 0$ for $\forall (i,j)$.} 
\end{array}
 \right\}.
\end{align*}
A divisor class $\Lambda \in M \subset {\rm Pic}(X)$
is represented by an exceptional curve because 
$\Lambda^2=( \Lambda | K_X )=-1$.
Moreover,
each class $\Lambda = d h-\sum_{i,m} \mu_i^m e_i^m \in M$ corresponds 
to (the proper transform of)
a unique plane curve $C_\Lambda$
passing through $P_i^m$ with multiplicity $\mu_i^m$
as its representative.
Let $F_\Lambda=F_\Lambda({\boldsymbol a};{\boldsymbol x})$ be the {\it normalized} defining polynomial
of $C_\Lambda$
such that 
the following key condition is satisfied:
\begin{equation} \label{eq:nor}
F_\Lambda({\boldsymbol a};1,1,1) =1.
\end{equation}
Notice that
$C_\Lambda$ never passes through the point
$P_0=[1:1:1]=L_1 \cap L_2 \cap L_3$
by definition.
Let us take for example $\Lambda=h-e_i^m-e_j^n \in M$,
which is represented by the line passing through
 $P_i^m$ and $P_j^n$;
thus we have
\[
F_{h-e_i^m-e_j^n}({\boldsymbol a};{\boldsymbol x})=
(1+b_i^m+b_j^n)x_\ell
-b_i^m x_m-b_j^n x_n 
\quad  \text{for}  \quad  \{\ell,m,n\}=\{1,2,3\}.
\]
We introduce  new variables $\tau_i^m$ attached to the points $P_i^m$, and
prepare a field $L=K({\boldsymbol x};{\boldsymbol \tau})$
of  rational functions in indeterminates  
$x_m$ and $\tau_i^m$ ($m=1,2,3$; $1 \leq i \leq\ell_m$)
with coefficient field $K={\mathbb C}({\boldsymbol a})$.
By means of  the normalized defining polynomial, 
we define a function $\tau: M \to L={\mathbb C}({\boldsymbol a})({\boldsymbol x};{\boldsymbol \tau})$,
called the {\it tau-function},
by the formula
\begin{equation}  \label{eq:deftau}
\tau(\Lambda)\prod_{i,m}\tau(e_i^m)^{\mu_i^m}=F_\Lambda({\boldsymbol a}; {\boldsymbol x})
\end{equation}
for $\Lambda=d h-\sum_{i,m} \mu_i^m e_i^m\in M$ $(d > 0)$,
and 
\begin{equation}
\tau(e_i^m)=\tau_i^m. 
\end{equation}

By imposing the following assumption:
\begin{equation} \label{eq:ass} 
w. \tau(\Lambda)=\tau(w.\Lambda),
\end{equation}
one can fix the action of Weyl group $W$ on $L={\mathbb C}({\boldsymbol a})({\boldsymbol x};{\boldsymbol \tau})$ in the following manner.
For a rational function $\varphi({\boldsymbol a};{\boldsymbol x};{\boldsymbol \tau}) \in L$,
we suppose that an element $w \in W$ acts as
\[
w .\varphi({\boldsymbol a};{\boldsymbol x};{\boldsymbol \tau}) 
=\varphi({\boldsymbol a} . w ;{\boldsymbol x}. w ;{\boldsymbol \tau}. w),
\]
that is, 
$w$ acts on the independent variables from the right.
We now determine the action of the generators $s_0$ and $s_i^m$ from (\ref{eq:ass}) as a 
necessary condition.
In view of $s_0(e_1^k)=h-e_1^i-e_1^j$ $(\{i,j,k\}=\{1,2,3\})$,
we have
\[
s_0(\tau_1^k)=\frac{F_{h-e_1^i-e_1^j}({\boldsymbol a};{\boldsymbol x})}{\tau_1^i\tau_1^j}.
\]
The other $\tau_i^m$'s 
$(i \neq 1)$
are kept still by $s_0$-action.
Applying $s_0$ to $F_{h-e_1^i-e_1^j}({\boldsymbol a};{\boldsymbol x} ) = \tau(h-e_1^i-e_1^j)\tau_1^i\tau_1^j$, we have
\begin{align*}
\text{(LHS)}
&=s_0(F_{h-e_1^i-e_1^j}({\boldsymbol a};{\boldsymbol x} ) )
\\
&=F_{h-e_1^i-e_1^j}(s_0({\boldsymbol a});s_0({\boldsymbol x}) ) 
\\
&=(1+\tilde{b}_1^i+\tilde{b}_1^j)s_0(x_k)
-\tilde{b}_1^i s_0(x_i)-\tilde{b}_1^j s_0(x_j),
\\
\text{(RHS)}
&=s_0( \tau(h-e_1^i-e_1^j)\tau_1^i\tau_1^j)
\\
&=\tau_1^k \tau(h-e_1^j-e_1^k)\tau(h-e_1^i-e_1^k)
\\
&=\frac{  F_{h-e_1^j-e_1^k}({\boldsymbol a};{\boldsymbol x} ) F_{h-e_1^i-e_1^k}({\boldsymbol a};{\boldsymbol x} )   }{\tau_1^1\tau_1^2\tau_1^3},
\end{align*}
where $\{i,j,k\}=\{1,2,3\}$ and 
\begin{equation}
\tilde{b}_1^i :=s_0(b_1^i) .
\end{equation}
Solving the above equations with respect to $s_0(x_i)$, we obtain
\[
s_0(x_i)=
\frac{(1+ \tilde{b}_1^i) f_jf_k+ \tilde{b}_1^j f_if_k+\tilde{b}_1^k f_if_j}{( 1+ \tilde{b}_1^1 +\tilde{b}_1^2 +\tilde{b}_1^3 )\tau_1^1\tau_1^2\tau_1^3},
\]
where 
\begin{equation}
f_i :=F_{h-e_1^j-e_1^k}({\boldsymbol a};{\boldsymbol x})
=(1+b_1^j+b_1^k)x_i-b_1^j x_j-b_1^k x_k
\quad
\text{for}  \quad  \{i,j,k\}=\{1,2,3\}.
\end{equation}
The action of $s_i^m$ is much simpler and is realized as (just a permutation of $\tau$-variables)    
\[ s_i^m(\tau_{ \{i,i+1 \}}^m) = \tau_{ \{i+1,i \}}^m,  \quad s_i^m({\boldsymbol x})= {\boldsymbol x}.\]    
Summarizing above, 
we have  the

\begin{thm} \label{thm:birat}
{\rm (I)}
Define
the birational transformations $s_0$ and $s_i^m$
by
\begin{subequations} \label{subeq:birat}
\begin{align}
s_0(\tau_1^i)&=\frac{f_i}{\tau_1^j \tau_1^k}, 
\label{eq:birat-a}
\\
s_0(x_i)
&=\frac{(1+ \tilde{b}_1^i) f_jf_k+ \tilde{b}_1^j f_if_k+\tilde{b}_1^k f_if_j}{( 1+ \tilde{b}_1^1 +\tilde{b}_1^2 +\tilde{b}_1^3 )\tau_1^1\tau_1^2\tau_1^3},
\label{eq:birat-b}
\end{align}
where $\{i,j,k\}=\{1,2,3\}$, 
and 
\begin{equation}  \label{eq:birat-c}
s_i^m(\tau_{ \{i,i+1 \}}^m) = \tau_{ \{i+1,i \}}^m.
\end{equation}
\end{subequations}
Then {\rm(\ref{subeq:birat})}  
 with {\rm (\ref{eq:act-on-a})} realize the Weyl group 
$W=W(T_{\ell_1,\ell_2,\ell_3})$
over the field ${\mathbb C}({\boldsymbol a})({\boldsymbol x};{\boldsymbol \tau})$.
\\
{\rm(II)}
 Moreover {\rm(\ref{eq:deftau})} and {\rm(\ref{eq:ass})} are consistent.
\end{thm}

\pf
(I) By direct computation, we can verify that (\ref{subeq:birat}) 
indeed satisfy the fundamental relations of $W$.

(II)
We will show that the formula (\ref{eq:deftau}) 
can be recovered inductively 
by (\ref{eq:ass}) and (\ref{subeq:birat}). 
First, if $\Lambda= h-e_1^i-e_1^j\in M$, then
(\ref{eq:deftau}) follows immediately from  (\ref{eq:birat-a}).
Next we assume (\ref{eq:deftau}) is true for $\Lambda \in M$.
It is enough to verify for $\Lambda'=w(\Lambda)$ where $w \in \{ s_0, s_i^m \}$ is a generator of $W$. 
Since the action of $s_i^m$ is just a permutation of $\tau$-variables (see (\ref{eq:birat-c})), 
we will concentrate our attention on  the only nontrivial case $w=s_0$.
Applying $s_0$ to  
(\ref{eq:deftau}),
we have
\begin{align*}
s_0(F_\Lambda({\boldsymbol a};{\boldsymbol x}))
&=
s_0\left(\tau(\Lambda)\prod_{i,j}(\tau_j^i)^{\mu_j^i}\right)
\\
&=\tau(s_0.\Lambda) \prod_{i,j}(\tau_j^i)^{\mu_j^i}
\prod_{i=1,2,3} \left(  \frac{s_0( \tau_1^i)}{\tau_1^i} \right)^{\mu_1^i}
\\
&=\tau(s_0.\Lambda) \prod_{i,j}(\tau_j^i)^{\mu_j^i}
\prod_{i=1,2,3} \left(   \frac{f_i}{\tau_1^1\tau_1^2\tau_1^3} \right)^{\mu_1^i}.
\end{align*}
Noticing 
$s_0.\Lambda=\Lambda+(\Lambda| \alpha_0)\alpha_0=\Lambda+(d-\mu_1^1-\mu_1^2-\mu_1^3)(h-e_1^1-e_1^2-e_1^3)$,
we can verify  
(\ref{eq:deftau}) for $\Lambda'=s_0. \Lambda$ 
immediately from the lemma below. 
\qed
\\

\begin{lemma}
\rm 
We have
\begin{equation}
s_0(F_{\Lambda}({\boldsymbol a};{\boldsymbol x}))
=\frac{F_{s_0. \Lambda} ({\boldsymbol a};{\boldsymbol x}) \prod_{i=1,2,3} (f_i)^{\mu_1^i} }{(\tau_1^1\tau_1^2\tau_1^3)^d}
\end{equation}
for $\Lambda = d h-\sum_{i,j} \mu_j^i e_j^i\in M$.
\end{lemma}

\pf
The multiplicity of the curve $C_\Lambda=\{F_\Lambda=0\}$
at $P_1^i=\{f_j=f_k=0\}$ $(\{i,j,k\}=\{1,2,3\})$
is  given by 
$\mu_1^i={\rm ord}_{P_1^i}(F_\Lambda)$.
That is, $F_\Lambda$ can be expressed as a polynomial in $f_i$ $(i=1,2,3)$ of the form
\[
F_\Lambda=\sum_{k_1+k_2+k_3=d} A_{k_1,k_2,k_3} (f_1)^{k_1} (f_2)^{k_2} (f_3)^{k_3},
\quad 
A_{k_1,k_2,k_3} \in {\mathbb C}({\boldsymbol a})
\]
such that $A_{k_1,k_2,k_3}=0$ unless $0 \leq k_i \leq d- \mu_1^i$ $(i=1,2,3)$.
By using 
$s_0(f_i)={f_jf_k}/{(\tau_1^1\tau_1^2\tau_1^3)}$,
we have
\begin{align*}
s_0(F_\Lambda) \frac{(\tau_1^1\tau_1^2\tau_1^3)^d}{\prod_{i=1,2,3} (f_i)^{\mu_1^i} }
&=
\sum_{k_1+k_2+k_3=d} A_{k_1,k_2,k_3} (f_1)^{d-\mu_1^1-k_1} (f_2)^{d-\mu_1^2-k_2} (f_3)^{d-\mu_1^3-k_3} =: G({\boldsymbol x}).
\end{align*}
We let
$\tilde{k}_i = d-\mu_1^i-k_i$, $\tilde{d}=2d-\mu_1^1-\mu_1^2-\mu_1^3$ and $\widetilde{A}_{\tilde{k}_1,\tilde{k}_2,\tilde{k}_3}=A_{k_1,k_2,k_3}$.
The polynomial $G({\boldsymbol x})$ can be then written as
\[
G({\boldsymbol x})= \sum_{\tilde{k}_1+\tilde{k}_2+\tilde{k}_3=\tilde{d}}\widetilde{A}_{\tilde{k}_1,\tilde{k}_2,\tilde{k}_3} (f_1)^{\tilde{k}_1} (f_2)^{\tilde{k}_2} (f_3)^{\tilde{k}_3},
\]
where $\widetilde{A}_{\tilde{k}_1,\tilde{k}_2,\tilde{k}_3}=0$ unless
$0 \leq \tilde{k}_i \leq d- \mu_1^i  = \tilde{d}-(d-\mu_1^j-\mu_1^k)$.
Therefore, 
the curve
$\{G({\boldsymbol x})=0\}$ represents
the divisor class $s_0.\Lambda$. 
Since $f_i|_{{\boldsymbol x}=(1,1,1)}=1$, 
the normalizing condition (\ref{eq:nor})
of $F_\Lambda$ yields 
$\sum A_{k_1,k_2,k_3}=1$.
Hence 
$G(1,1,1)=\sum \widetilde{A}_{\tilde{k}_1,\tilde{k}_2,\tilde{k}_3} 
=\sum A_{k_1,k_2,k_3}=1$,
and so $G=F_{s_0. \Lambda}$ as desired.
\qed 
\\

\begin{remark}\rm
By virtue of  the geometric characterization (\ref{eq:deftau}) of  tau-function, 
we can trace the resulting value for any iteration of the Weyl group action 
(Theorem~\ref{thm:birat})
by using only the defining polynomials of appropriate plane curves,
although it is in general  difficult to compute composition of given rational maps.
Also, note that the defining polynomial of a curve
has a determinantal expression 
involving information with respect to the multiplicities at 
enough points; see e.g. \cite{kmnoy06}.
\end{remark}

\section{Affine case and difference Painlev\'e equations of type E}
\label{sect:aff}

From Theorem~\ref{thm:birat},
one can also realize the Weyl group 
$W=W(T_{\ell_1,\ell_2,\ell_3})$ 
as 
birational transformations of inhomogeneous coordinates of ${\mathbb P}^2$.
In this context,  tau-functions play roles of {\it heights}
in the sense that the original inhomogeneous coordinate is recovered as a ratio of them.
In this section, we consider the affine case: $E_6^{(1)}=T_{3,3,3}$, 
$E_7^{(1)}=T_{4,4,2}$ and
$E_8^{(1)}=T_{6,3,2}$,
in which our realization of Weyl groups on inhomogeneous coordinates 
is equivalent to that of Sakai's formulation \cite{sak01}, 
and therefore it is relevant to discrete Painlev\'e equations.
Recall that discrete Painlev\'e equations are divided into three types:
difference, $q$-difference or elliptic-difference; our setting corresponds to the difference one.
For each case, we will derive
the {\it difference Painlev\'e equations} 
in terms of 
appropriately chosen coordinates.

\subsection{$E_6^{(1)}$ case ($\ell_1=\ell_2=\ell_3=3$)}

The Dynkin diagram of type $E_6^{(1)}=T_{3,3,3}$ is
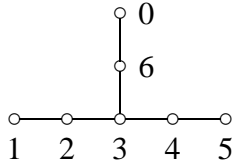
\begin{center}
\begin{picture}(200,70)

\put(60,20){\circle{4}}    \put(57,5){$1$}
\put(78,20){\line(-1,0){16}}
\put(80,20){\circle{4}}    \put(77,5){$2$}
\put(98,20){\line(-1,0){16}}
\put(100,20){\circle{4}}    \put(97,5){$3$}
\put(102,20){\line(1,0){16}}
\put(120,20){\circle{4}}    \put(117,5){$4$}
\put(122,20){\line(1,0){16}}
\put(140,20){\circle{4}}    \put(137,5){$5$}

\put(100,22){\line(0,1){16}}
\put(100,40){\circle{4}}   \put(107,36){$6$}
\put(100,42){\line(0,1){16}}
\put(100,60){\circle{4}}   \put(107,56){$0$}

\end{picture}
\end{center}
Let $(a_0,a_1,a_2,a_3,a_4,a_5,a_6)$ be the root variables of $E_6^{(1)}$.
The transformation from the original variables $a_0, a_i^m$ 
(used in Section~\ref{sect:bir})
is given as follows:
\[
a_0 : \Rightarrow  a_3, \quad
a^1_{\{1,2 \}}   : \Rightarrow a_{\{2,1\}}, \quad
 a^2_{\{1,2 \}}   : \Rightarrow a_{\{4,5\}}, \quad
 a^3_{\{1,2 \}}   : \Rightarrow a_{\{6,0\}}.
\]
We define the action of simple reflections $s_i$ on the root variables by $s_i(a_j)=a_j-a_iC_{ij}$, where $C_{ij}$ is the Cartan matrix of $E_6^{(1)}$.
Let $\iota_i$ $(i=1,2,3)$ be the diagram automorphisms defined by 
$\iota_1(a_{\{0,1,2,3,4,5,6\}})=a_{\{5,1,2,3,6,0,4\}}$,
$\iota_2(a_{\{0,1,2,3,4,5,6\}})=a_{\{1,0,6,3,4,5,2\}}$
and
$\iota_3(a_{\{0,1,2,3,4,5,6\}})=a_{\{0,5,4,3,2,1,6\}}$.

Let us consider the change of the variables
$\varphi:{\mathbb P}^2 \to {\mathbb P}^1 \times {\mathbb P}^1$
given by 
\[\varphi:[x_1:x_2:x_3] \mapsto (u,v)=\left( \frac{F_{h-e^1_1-e^2_1}({\boldsymbol a}; {\boldsymbol x})}{x_2-x_3},\frac{F_{h-e^1_1-e^2_1}({\boldsymbol a}; {\boldsymbol x})}{x_1-x_3}\right).
\]
This transformation is obtained by 
blowing up the points $P^1_1$ and $P^2_1$,
and  blowing down (the proper transform of) 
the line $\{ F_{h-e^1_1-e^2_1}= 0\}$
which passes
through the two points.
Note that the sum of three lines $L_1 \cup L_2 \cup L_3$, 
which is invariant under the action of the Weyl group,
is generically transformed into 
$\{ u= \infty\} \cup \{v=\infty\} \cup \{u=v\}$
by $\varphi$.

From Theorem~\ref{thm:birat},
we have the following realization of 
$\widetilde{W}(E_6^{(1)})= \langle s_0, \ldots, s_6, \iota_1, \iota_2,\iota_3 \rangle$ 
on $(u,v)$:
\begin{align*}
&
s_2(u)=\frac{u(v-a_2)}{v}, \quad s_2(v)=v-a_2, \\
&
s_3(u)=u+a_3, \quad s_3(v)=v+a_3, \\
&
s_4(u)=u-a_4, \quad s_4(v)=\frac{v(u-a_4)}{u}, \\
&
\iota_1(u)=-u-a_3, \quad \iota_1(v)= \frac{v(u+a_3)}{u-v}, \\
&
\iota_2(u)= \frac{u(v+a_3)}{v-u}, \quad \iota_2(v)=-v-a_3, \\
&
\iota_3(u)=v, \quad \iota_3(v)=u.
\end{align*}
Here the symbol 
$\widetilde W$ stands for the extended group of $W$ 
by its diagram automorphisms.
The birational action of a translation in $\widetilde{W}(E_6^{(1)})$
yields the difference Painlev\'e equation.
We take an element 
\[T=(\iota_3\iota_1s_6s_0s_3s_4s_6s_3s_5s_4)^2
\]
acting on the root variables as their shifts:
$T(a_0,a_1,a_2,a_3,a_4,a_5,a_6)=(a_0,a_1,a_2+\delta,a_3,a_4-\delta,a_5,a_6)$,
where 
$\delta=a_0+a_1+2a_2+3a_3+2a_4+a_5+2a_6$
corresponds to  the null root.
Let $\overline{u}=T(u)$ and $\underline{v}=T^{-1}(v)$.
Then we have the system of functional equations:
\begin{subequations}
\begin{align}
(\overline{u}-v)(u-v)
&=
\frac{v(v+a_3)(v+a_3+a_6)(v+a_3+a_6+a_0)}{(v-a_2)(v-a_1-a_2)},
\\
(\underline{v}-u)(v-u)
&=
\frac{u(u+a_3)(u+a_3+a_6)(u+a_3+a_6+a_0)}{(u-a_4)(u-a_4-a_5)}.
\end{align}
\end{subequations}
This is called the {\it difference Painlev\'e equation of type $E_6^{(1)}$}.

\subsection{$E_7^{(1)}$ case ($\ell_1=\ell_2=4, \ell_3=2$)}

The Dynkin diagram of type $E_7^{(1)}=T_{4,4,2}$ is
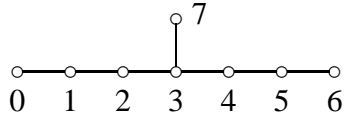
\begin{center}
\begin{picture}(200,50)

\put(40,20){\circle{4}}    \put(37,5){$0$}
\put(58,20){\line(-1,0){16}}
\put(60,20){\circle{4}}    \put(57,5){$1$}
\put(78,20){\line(-1,0){16}}
\put(80,20){\circle{4}}    \put(77,5){$2$}
\put(98,20){\line(-1,0){16}}
\put(100,20){\circle{4}}    \put(97,5){$3$}
\put(102,20){\line(1,0){16}}
\put(120,20){\circle{4}}    \put(117,5){$4$}
\put(122,20){\line(1,0){16}}
\put(140,20){\circle{4}}    \put(137,5){$5$}
\put(142,20){\line(1,0){16}}
\put(160,20){\circle{4}}    \put(157,5){$6$}

\put(100,22){\line(0,1){16}}
\put(100,40){\circle{4}}   \put(106,38){$7$}

\end{picture}
\end{center}
Let $(a_0,a_1,a_2,a_3,a_4,a_5,a_6,a_7)$ be the root variables of $E_7^{(1)}$.
Here we have replaced 
the variables
$a_0, a_i^m$ (used in Section~\ref{sect:bir})
as
\[
a_0  : \Rightarrow a_3, \quad
a^1_{\{1,2,3 \}}  : \Rightarrow a_{\{2,1,0\}}, \quad
 a^2_{\{1,2,3 \}}   : \Rightarrow a_{\{4,5,6\}}, \quad
 a^3_{1}   : \Rightarrow a_7.
\]
Define the action of simple reflections $s_i$ on the root variables by $s_i(a_j)=a_j-a_iC_{ij}$, where $C_{ij}$ is the Cartan matrix of $E_7^{(1)}$.
Let $\iota$ be the diagram automorphism defined by 
$\iota(a_{\{0,1,2,3,4,5,6,7\}})=a_{\{6,5,4,3,2,1,0,7\}}$.

We first blow up 
the two points $P^3_1$, $P^3_2$
and blow down (the proper transform of) the line 
$L_3=\{ x_1=x_2 \}$.
This procedure yields
the birational map
$\varphi_1: {\mathbb P}^2 \to {\mathbb P}^1 \times {\mathbb P}^1$ 
given by
\[
\varphi_1: [x_1:x_2:x_3] \mapsto (u_1,v_1)
= \left( \frac{ (b_1^3+1)(x_1+x_2)-2 b_1^3 x_3 }{x_1-x_2},
 \frac{ (b_2^3+1)(x_1+x_2)-2 b_2^3 x_3 }{x_1-x_2} \right).
\]
In addition, 
we apply 
\[
\varphi_2: (u_1,v_1) \mapsto 
(u,v)=\left(u_1+\frac{a_0+2a_1+3a_2-3a_4-2a_5-a_6}{4},
v_1+\frac{a_0+2a_1+3a_2-3a_4-2a_5-a_6}{4}
\right).
\]
Note that $L_1 \cup L_2 \cup L_3$ is generically transformed by $\varphi= \varphi_2 \circ \varphi_1$
into 
two curves $\{u-v+a_7=0\} \cup \{u-v-a_7=0\} \subset  {\mathbb P}^1 \times {\mathbb P}^1$
of bidegree $(1,1)$
tangent to each other at $(\infty,\infty)$.

From Theorem~\ref{thm:birat},
we have the following realization of $\widetilde{W}(E_7^{(1)})$ on $(u,v)$:
\begin{align*}
&
s_2(u)=u-a_2, \quad s_2(v)=v-a_2, \quad
s_3(v)= \tilde{v}, \quad 
s_4(u)=u+a_4, \quad s_4(v)=v+a_4, \\
&
\iota(u)=-u, \quad \iota(v)=-v,
\end{align*}
where $\tilde{v}=\tilde{v}(u,v)$ is a rational function determined by 
\[
\frac{u-\tilde{v}+a_3+a_7}{u-\tilde{v}-a_3-a_7}
=\frac{(u+a_3)(u-v+a_7)}{(u-a_3)(u-v-a_7)}.
\]
Let us take an element
\[
T=(s_7s_3s_2s_1s_0s_4s_5s_6s_3s_2s_1s_4s_5s_3s_2s_4s_3)^2
\in W(E_7^{(1)})
\]
acting on the root variables as 
$T(a_0,a_1,a_2,a_3,a_4,a_5,a_6,a_7)=(a_0,a_1,a_2,a_3-\delta,a_4,a_5,a_6,a_7+2 \delta)$,
where
$\delta=a_0+2a_1+3a_2+4a_3+3a_4+2a_5+a_6+2a_7$.
Let $\overline{u}=T(u)$ and $\underline{v}=T^{-1}(v)$.
Then we have the system of functional equations:
\begin{subequations}
\begin{align}
\lefteqn{\frac{(\underline{v}-u+a_7-\delta)(v-u+a_7)}{(\underline{v}-u-a_7+\delta)(v-u-a_7)}}
\\
& \qquad =
\frac{(u+a_3)(u+a_3+2a_4)(u+a_3+2a_4+2 a_5)(u+a_3+2a_4+2 a_5+2a_6)}{(u-a_3)(u-a_3-2a_2)(u-a_3-2a_2-2a_1)(u-a_3-2a_2-2a_1-2a_0)},
\nonumber
\\
\lefteqn{\frac{(\overline{u}-v-a_7-\delta)(u-v-a_7)}{(\overline{u}-v+a_7+\delta)(u-v+a_7)}}
\\
& \qquad =
\frac{(v+a_3+a_7)(v+a_3+2a_4+a_7)(v+a_3+2a_4+2a_5+a_7)(v+a_3+2a_4+2a_5+2a_6+a_7)}{(v-a_3-a_7)(v-a_3-2a_2-a_7)(v-a_3-2a_2-2a_1-a_7)(v-a_3-2a_2-2a_1-2a_0-a_7)}.
\nonumber
\end{align}
\end{subequations}
This is called the {\it difference Painlev\'e equation of type $E_7^{(1)}$}.

\subsection{$E_8^{(1)}$ case ($\ell_1=6, \ell_2=3, \ell_3=2$)}

The Dynkin diagram of type $E_8^{(1)}=T_{6,3,2}$ is
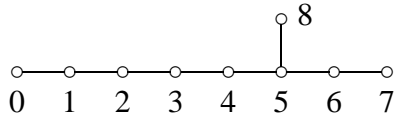
\begin{center}
\begin{picture}(220,50)

\put(40,20){\circle{4}}    \put(37,5){$0$}
\put(58,20){\line(-1,0){16}}
\put(60,20){\circle{4}}    \put(57,5){$1$}
\put(78,20){\line(-1,0){16}}
\put(80,20){\circle{4}}    \put(77,5){$2$}
\put(98,20){\line(-1,0){16}}
\put(100,20){\circle{4}}    \put(97,5){$3$}
\put(102,20){\line(1,0){16}}
\put(120,20){\circle{4}}    \put(117,5){$4$}
\put(122,20){\line(1,0){16}}
\put(140,20){\circle{4}}    \put(137,5){$5$}
\put(142,20){\line(1,0){16}}
\put(160,20){\circle{4}}    \put(157,5){$6$}

\put(162,20){\line(1,0){16}}
\put(180,20){\circle{4}}    \put(177,5){$7$}

\put(140,22){\line(0,1){16}}
\put(140,40){\circle{4}}   \put(146,38){$8$}

\end{picture}
\end{center}
Let $(a_0,a_1,a_2,a_3,a_4,a_5,a_6,a_7,a_8)$ be the root variables of $E_8^{(1)}$.
The interpretation from the original variables
$a_0, a_i^m$ (in Section~\ref{sect:bir})
is given by
\[
a_0 : \Rightarrow a_5, \quad
a^1_{\{1,2,3,4,5 \}}  : \Rightarrow a_{\{4,3,2,1,0\}}, \quad
 a^2_{\{1,2 \}}   : \Rightarrow a_{\{6,7\}}, \quad
 a^3_{1}  : \Rightarrow a_8.
\]
Define the action of simple reflections $s_i$ on the root variables by $s_i(a_j)=a_j-a_iC_{ij}$, where $C_{ij}$ is the Cartan matrix of $E_8^{(1)}$.

First  we blow up the two points $P^2_1$, $P^3_1$  
and blow down 
(the proper transform) of the line 
$\{ F_{h-e^2_1-e^3_1}=0\}$.
This yields the birational map
$\varphi_1:{\mathbb P}^2 \to {\mathbb P}^1 \times {\mathbb P}^1$ of the form
\[
\varphi_1:[x_1:x_2:x_3] \mapsto (u_1,v_1)=
\left( 
\frac{F_{h-e^2_1-e^3_1}({\boldsymbol a}; {\boldsymbol x}) }{x_1-x_2},
\frac{F_{h-e^2_1-e^3_1}({\boldsymbol a}; {\boldsymbol x}) }{x_1-x_3}
 \right).
\]
Secondly we blow up $P^2_2: (u_1,v_1)=(-a_6,\infty)$ and $P^3_2:(u_1,v_1)=(\infty, -a_8)$,
and blow down  
the lines $\{ u_1= -a_6\}$ and $\{ u_1=\infty\}$.
Then we have the birational map   
$\varphi_2:{\mathbb P}^1 \times {\mathbb P}^1\circlearrowleft$
written as 
\[
\varphi_2:(u_1,v_1) \mapsto (u_2,v_2)=(u_1+a_6,(u_1+a_6)(v_1+a_8)).
\]
Thirdly,
by blowing up $(u_2,v_2)=(0,0)$ and $P^2_3:(u_2,v_2)=(-a_7,\infty)$ and contracting $\{v_2=0\}$ and $\{v_2=\infty\}$,
we obtain 
the birational map
$\varphi_3:{\mathbb P}^1 \times {\mathbb P}^1 \circlearrowleft$,
\[
\varphi_3:(u_2,v_2) \mapsto (u_3,v_3)=\left(\frac{v_2(u_2+a_7)}{u_2},v_2 \right).
\]
Finally we apply the linear transformation
\[
\varphi_4:(u_3,v_3) \mapsto 
(u,v)=\left(4 u_3+(a_6+a_7-a_8)^2, 4 v_3+ (a_6-a_8)^2 \right).
\]
We see that $L_1 \cup L_2 \cup L_3$ is generically transformed by $\varphi=\varphi_4 \circ \varphi_3 \circ \varphi_2 \circ \varphi_1$
into the curve
$\{ (u-v)^2-2 {a_7}^2(u+v)+{a_7}^4=0 \}
\subset
{\mathbb P}^1 \times {\mathbb P}^1$
of bidegree $(2,2)$
with a cusp at $(\infty, \infty)$.

By virtue of Theorem~\ref{thm:birat},
we have
the following  realization of $W(E_8^{(1)})$ on $(u,v)$:
\[
s_6(u)=\tilde{u}, \quad
s_7(u)=v, \quad s_7(v)=u,
\]
where $\tilde{u}=\tilde{u}(u,v)$ is a rational function determined by 
\[
\frac{\tilde{u}-(a_7+a_8)^2}{\tilde{u}-(a_7-a_8)^2}=
\frac{u-(a_6+a_7+a_8)^2}{u-(a_6+a_7-a_8)^2}
\cdot
\frac{v-(a_6-a_8)^2}{v-(a_6+a_8)^2}.
\]
Alternatively, with respect to the variable (cf. \cite{msy03})
\[\Gamma=\frac{v-u+{a_7}^2}{2a_7},\]
we can describe simply the action of $s_6$ as a
linear fractional transformation
\[
s_6(\Gamma)= \frac{(v+(a_6-a_8)(a_6+a_8))\Gamma + 2 a_6 v}{2 a_6 \Gamma + v+(a_6-a_8)(a_6+a_8)}.
\]
Consider an element
\[
T=
(s_6s_5s_4s_3s_2s_1s_0s_8s_5s_4s_3s_2s_1
s_6s_5s_4s_3s_2s_8s_5s_4s_3
s_6s_5s_4s_8s_5
s_6s_7)^2
\in W(E_8^{(1)})
\]
acting on the root variables as 
$T(a_0,a_1,a_2,a_3,a_4,a_5,a_6,a_7,a_8)=(a_0,a_1,a_2,a_3,a_4,a_5,a_6+\delta,a_7-2 \delta,a_8)$,
where
$\delta=a_0+2a_1+3a_2+4a_3+5a_4+6a_5+4a_6+2a_7+3a_8$.
Let $\overline{u}=T(u)$ and $\underline{v}=T^{-1}(v)$.
Then we have the system of functional equations:
\begin{subequations}
\begin{align}
\frac{\left(\overline{u}-v-(a_7-\delta)^2 \right)\left(  u-v-{a_7}^2\right) + 4 a_7(a_7-\delta)v}{ 2 a_7\left(\overline{u}-v-(a_7-\delta)^2 \right)+2(a_7-\delta)\left(  u-v-{a_7}^2\right)}
&= \frac{G_4({\boldsymbol \theta};v)}{G_3({\boldsymbol \theta};v)},
\\
\frac{\left(u-\underline{v}+(a_7+\delta)^2 \right) \left(  u-v+{a_7}^2\right) + 4 a_7(a_7+\delta)u}{ 2 a_7\left(u-\underline{v}+(a_7+\delta)^2 \right)+2(a_7+\delta)\left(  u-v+{a_7}^2\right)}
&= \frac{G_4({\boldsymbol \theta}+a_7  {\boldsymbol 1};u)}{G_3({\boldsymbol \theta}+a_7  {\boldsymbol 1};u)},
\end{align}
\end{subequations}
called the
{\it difference Painlev\'e equation of type $E_8^{(1)}$}.
Here we introduce the polynomials
\begin{align*}
G_4({\boldsymbol \theta};v)
&=v^4+\sigma_2 v^3+\sigma_4 v^2+\sigma_6 v+\sigma_8,
\\
G_3({\boldsymbol \theta};v)
&=\sigma_1 v^3+\sigma_3 v^2+\sigma_5 v+\sigma_7,
\end{align*}
where
$\sigma_i$ is the $i$-th elementary  symmetric function of 
the eight variables
${\boldsymbol \theta}=(\theta_1,\ldots, \theta_8)$
with
 $\theta_1=a_6-a_8$, $\theta_2=s_8(\theta_1)(=a_6+a_8)$,  
  $\theta_3=s_5(\theta_2)$, $\theta_4=s_4(\theta_3)$, $\theta_5=s_3(\theta_4)$,
$\theta_6=s_2(\theta_5)$, $\theta_7=s_1(\theta_6)$ 
and
 $\theta_8=s_0(\theta_7)$,
and the symbol ${\boldsymbol 1}$ denotes $(1, \ldots,1)$.

\begin{remark} \rm
For reference, we mention some  known results about 
the difference Painlev\'e equations of type $E^{(1)}_r$
($r=6,7,8$);
these equations were originally discovered by Grammaticos-Ohta-Ramani
\cite{org01}
(see also  good review articles  \cite{gr, sak07} with a list of  discrete Painlev\'e equations). 
The associated Lax formalisms (or linear problems) were established by Arinkin-Borodin \cite{ab}  for $E_6^{(1)}$ and by Boalch \cite{boa} for  $E_7^{(1)}$ and $E_8^{(1)}$,
respectively.
It was recently reported by Kajiwara \cite{kaj} that
for all $E^{(1)}_r$ cases
they admit
special solutions in terms of the hypergeometric functions in Askey's scheme
as well as the $q$-difference ones \cite{kmnoy05}.

It often occurs 
in discrete/continuous  Painlev\'e equations  
that
an interesting class of solutions 
(written by means of Schur function and its variations)
is generated by the Weyl group action from 
a fixed point of diagram automorphisms.
In connection with integrable systems like KP hierarchy, 
we will study elsewhere such solutions based on 
the framework of tau-functions  
(cf. \cite{mt, tsu05}).
\end{remark}

\small
\noindent 
{\it Acknowledgements.} \
The author wishes to thank  
Kenji Kajiwara,Tetsu Masuda, Mikio Murata and Yasuhiko Yamada
for fruitful discussions.
Research of the author is supported in part by
JSPS Grant 19840039.

\end{document}